\documentclass[%
 aip,
 amsmath,amssymb,
 reprint,%
]{revtex4-1}

\usepackage{graphicx}
\usepackage{dcolumn}
\usepackage{bm}

\usepackage[utf8]{inputenc}
\usepackage[T1]{fontenc}
\usepackage{mathptmx}
\usepackage{etoolbox}

\usepackage{comment}
\usepackage{braket}
\usepackage{physics}

\usepackage{color}

\makeatletter
\def\@email#1#2{%
 \endgroup
 \patchcmd{\titleblock@produce}
  {\frontmatter@RRAPformat}
  {\frontmatter@RRAPformat{\produce@RRAP{*#1\href{mailto:#2}{#2}}}\frontmatter@RRAPformat}
  {}{}
}%
\makeatother
\begin{document}

\preprint{AIP/123-QED}

\title[Electronic and Spin States at Edges of Finite $p$-orbital Helical Atomic Chain]{Electronic and Spin States at Edges of Finite $p$-orbital Helical Atomic Chain}
\author{Takemitsu Kato}
 \email{tKato.take@gmail.com}
 \altaffiliation[]{Department of Physics Engineering, Faculty of Engineering, Mie University, Tsu, Mie, 514-8507, Japan}
\author{Yasuhiro Utsumi}%
 \email{utsumi@phen.mie-u.ac.jp}
\affiliation{
Department of Physics Engineering, Faculty of Engineering, Mie University
}

\author{Ora Entin-Wohlman}
\email{orawohlman@gmail.com}
 \altaffiliation[]{School of Physics and Astronomy, Tel Aviv University, Tel Aviv 6997801, Israel}
\author{Amnon Aharony}
\email{aaharonyaa@gmail.com}
\affiliation{%
School of Physics and Astronomy, Tel Aviv University
}

\date{\today}

\begin{abstract}
In connection to the chiral-induced spin-selectivity (CISS) effect, we theoretically analyze the electronic and spin states  of edges of a finite $p$-orbital helical atomic chain with the intra-atomic spin-orbit interaction (SOI).
This model can host the spin-filtering state in which two up spins propagate in one direction and two down spins propagate in the opposite direction without breaking the time-reversal symmetry.
The enhancement of charge modulations concentrated at the edges due to the evanescent states is induced, although the spin density is absent because of the time-reversal symmetry (TRS). 
A Zeeman field at an edge of the atomic chain, which breaks the TRS, yields a finite spin polarization, whose direction depends on the chirality of the molecule. 
The chirality change induces a reasonable amount of the energy difference, which may provide an insight into the enantioselective adsorption of chiral molecules on the ferromagnetic surface. 
\end{abstract}

\maketitle

\section{\label{intro}Introduction}

The chiral-induced spin selectivity (CISS) effect is a spin transport phenomenon that is specific to chiral materials:
When an electron is injected into chiral molecules such as DNA, the electron spin is selectively separated depending on the chirality.
This phenomenon exhibits a high spin polarization, and has attracted attention as a quantum phenomenon that occurs at room temperature.~\cite{Mishra}
Although the CISS effect has been observed in many experiments using various chiral materials, a fully convincing theoretical explanation has not yet been proposed.~\cite{Evers2022}

There is another interesting phenomenon associated to the CISS, the enantioselectivity: Molecules with a specific chirality selectively adsorbed to the surface of the magnetic substrate.~\cite{Banerjee,Naaman2020,Bloom2020}
Theoretically this effect is explained as the consequence of the spin-dependent dispersion force,~\cite{Naaman2020,Kumar2017} the charge redistribution of the molecule attached to substrate.~\cite{Fransson}
Reference~\onlinecite{Kishine} argues that the electric toroidal monopole is behind this phenomenon.
Another experiment suggests that the chiral molecule attached to the superconducting substrate induces the Shiba state.~\cite{Alpern2019}

\begin{figure}
\includegraphics[scale=0.23]{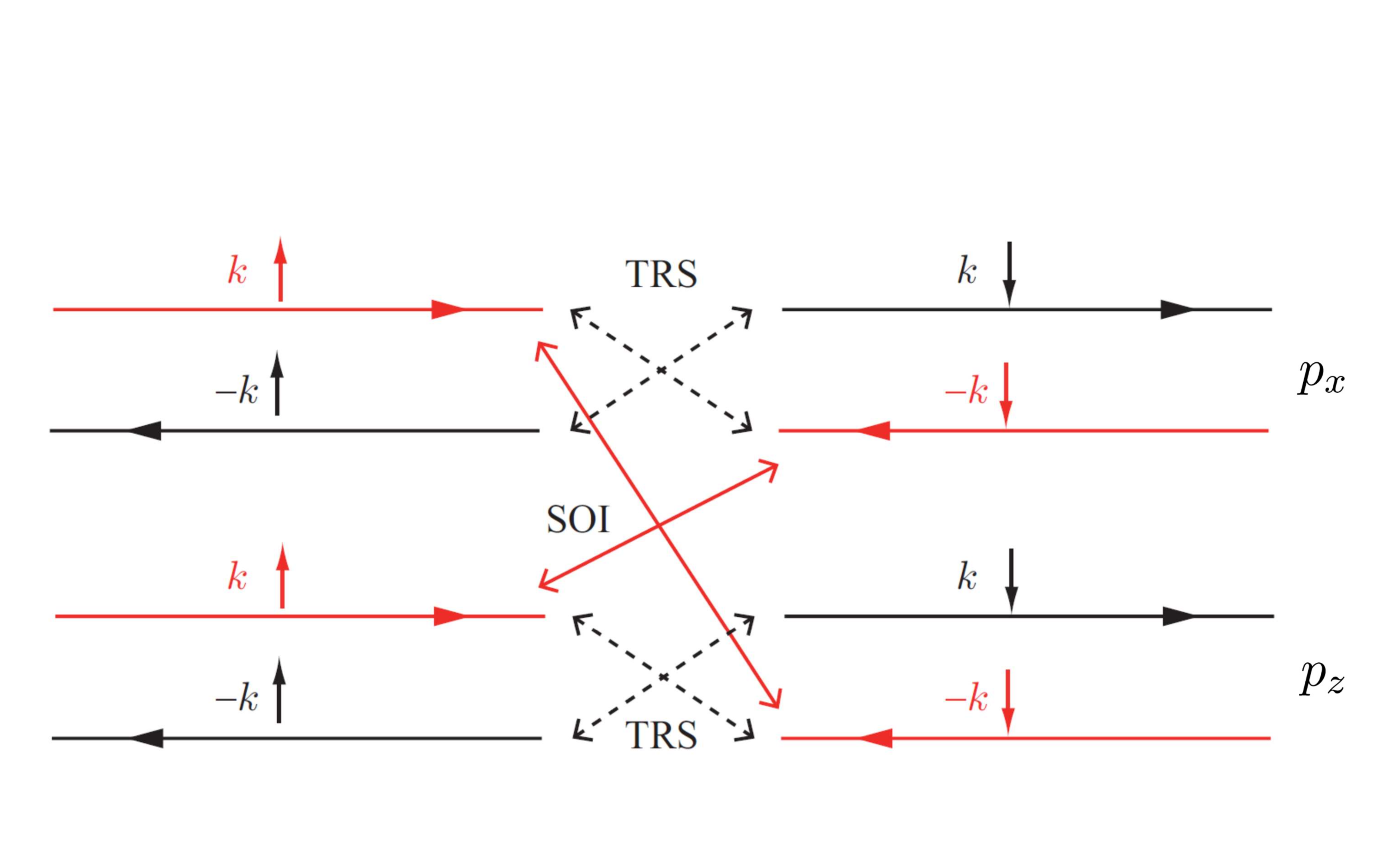}
\caption{\label{trs}
Schematic picture of the two-terminal and two-orbital spin filtering mechanism.
Different orbitals are mixed by the SOI, allowing the creation of opposite-spin states that flow in the opposite direction while maintaining time-reversal symmetry.}
\end{figure}

In the above mentioned works, the electronic and spin states at edges of the chiral molecule seem to be important.
In this study, we analyze this issue using a $p$-orbital helical atomic chain with the intra-atomic spin-orbit interaction (SOI), which is a toy model of helical molecules such as DNA.~\cite{YU2022,PhysRevB.102.035445}
Previously, we calculated the band structure of an infinite chain~\cite{YU2022} and the transmission probability of the molecular junction~\cite{PhysRevB.102.035445} to investigate the spin-filtering.
In this paper, we focus on the finite molecular chain to analyze the behavior of spin at the edges. 
In the presence of the strong crystal field along the tangential direction of the helix, this model is effectively reduced to the two-orbital channel one-dimensional chain with the SOI, which flips both the spin and orbital channels, see Fig.~\ref{trs}.
More explicitly, the effective Hamiltonian may be written as follows:
\begin{align}
V=&\ a\ c^\dag_{p_z;k\uparrow}c^{}_{p_x;-k\downarrow}-a\ c^\dag_{p_x;k\uparrow}c^{}_{p_z;-k\downarrow}\notag\\
&+a^*\ c^\dag_{p_x;-k\downarrow}c^{}_{p_z;k\uparrow}-a^*\ c^\dag_{p_z;-k\downarrow}c^{}_{p_x;k\uparrow}, \label{Heff}
\end{align}
where $a$ is a complex number.
Here $c^{}_{o;k \sigma_s}$ is an annihilation operator of an  electron with the orbital channel $o$ ($=p_x,p_z$), 
the wave number $k$ and the spin $\sigma_s$ ($=\uparrow, \downarrow$).
This Hamiltonian has the time-reversal symmetry (TRS), i.e. it is even under the time reversal, $\hat\Theta V \hat\Theta^{-1}=V$, where $\hat\Theta$ is the time-reversal operator.

The Hamiltonian Eq.~(\ref{Heff}) hybridyzes the left-going up-spin and the right-going down-spin and forms a standing wave.
As a consequence the right-going up-spin and the left-going down-spin form the helical state, which is responsible for the spin-filtering.
One can naively expect that, in a finite chain, the down-spin accumulates at one edge and the up-spin accumulates at the other edge.
We will discuss this issue in this paper.

The structure of the paper is the following:
In Sec.~\ref{porbit}, we introduce the $p$-orbital helical atomic chain and present out numerical results.
Furthermore, we present an analytic explanation of the electron state at the edges in Sec.~\ref{chargeA}.
In Sec.~\ref{magnet}, we introduce a Zeeman field at one edge, to imitate an chiral molecule adsorbed on a ferromagnetic substrate and discuss a connection to the enantioselectivity in CISS.
Section~\ref{conclusion} concludes our findings.

\section{\label{porbit}$\bm{p}$-orbital helical atomic chain}

\begin{figure}
\includegraphics[scale=0.2]{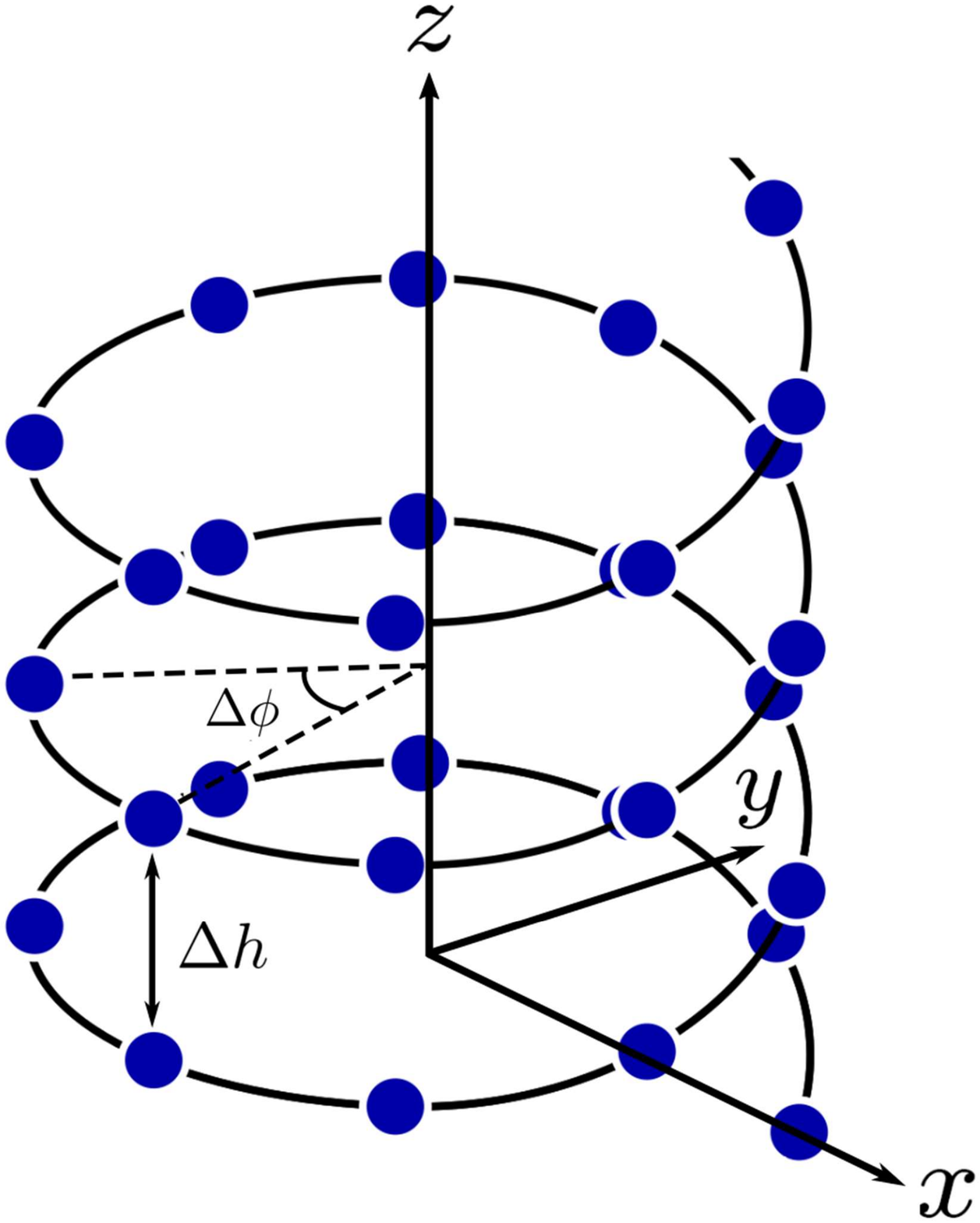}
\caption{\label{dna} Schematic picture of a helical atomic chain.
The model is a right-handed system in which the $z$-axis coincides with the helical axis.}
\end{figure}

The position of an atom on the helical atomic chain [Fig. \ref{dna}] is,
\begin{align}
{\bm R}(\phi_n)=[R\cos(\phi_n), R\sin(p\phi_n), \Delta h  \phi_n/(2\pi)].
\end{align}
where $R$, $n$, $\Delta h$ and $\phi_n=\Delta \phi \, n$ represent the radius, site number, the pitch, and the rotation angle around the z-axis, respectively. 
Here $p=+1 (-1)$ indicates the right (left)-handed helix.
The angle between the neigbouring atoms is $\Delta \phi=2\pi/N$, where $N$ is the number of atoms in a unit cell.
The helix is characterized by the normalized torsion $\tau$ and the curvature $\kappa=\sqrt{1-\tau^2}$:
\begin{align}
\tau=\frac{p\Delta h/(2\pi)}{\sqrt{R^2+[\Delta h/(2\pi)]^2}} .
\end{align}
The Hamiltonian of the helical atomic chain [Fig. \ref{dna}] is given by,
\begin{align}
H=\sum_{n=1}^{MN} \biggl( -&\tilde c^\dagger_{n+1}\bm J\otimes\sigma^{}_0\ \tilde c^{}_{n} +\mathrm{H.c.} \notag\\
&+\Delta_{\mathrm{so}} \tilde c^\dagger_{n}\ \bm L \cdot \bm \sigma\ \tilde c^{}_{n}\notag\\
&+K_t \tilde c^\dagger_{n} \left[\left(\bm t (\phi_n)\cdot\bm L\right)^2-\bm 1_3\right]\otimes\sigma^{}_0\ \tilde c^{}_{n} \biggl). \label{fullH}
\end{align}
The first line represents the electron hopping between nearest neighboring atoms.
The second line involves intra-atomic spin-orbit interactions (SOI), which are crucial for realizing the spin filter. 
The third line denotes the crystalline field, characterizing the helical structure of the chiral molecules.
Below, we provide detailed descriptions for each term.

$M$ is the number of unit cells and $\sigma_0$ represents a $2 \times 2$ unit matrix in spin space.
The $n$-th atom hosts $p$ orbitals and the vector of creation operators is
\begin{equation}
\tilde c^\dag_{n}=
[\tilde c^\dag_{n;x\uparrow} \;
\tilde c^\dag_{n;x\downarrow} \;
\tilde c^\dag_{n;y\uparrow} \;
\tilde c^\dag_{n;y\downarrow} \;
\tilde c^\dag_{n;z\uparrow} \;
\tilde c^\dag_{n;z\downarrow}],
\end{equation}
where $\tilde{c}^\dag_{n;o \sigma_s}$ ($o=p_x,p_y,p_z$) creates $\sigma_s$ spin at the $o$ orbitals in an atom at site $n$. 
This system satisfies the helical symmetry: 
The matrix ${\bm J}$ of the first term in the Hamiltonian is parameterized by three real numbers $J$, $\alpha$, and $\varphi$,~\cite{YU2022}
\begin{align}
{\bm J}=J
\begin{pmatrix}
\alpha \cos (p\varphi) &  -\alpha \sin (p\varphi)&0\\\alpha \sin (p\varphi)& \alpha \cos (p\varphi) &0\\0&0&1 \end{pmatrix}. \label{mJ}
\end{align}
In the following, we take $J>0$.
The second line is for the SOI, where ${\bm \sigma}=(\sigma_x,\sigma_y,\sigma_z)$ is the vector of Pauli matrices and
${\bm L}=(L_x,L_y,L_z)$ is the vector of the orbital angular momentum operators:
\begin{align}
L_x&=\begin{pmatrix}
0&0&0\\0&0&-i\\0&i&0
\end{pmatrix}\\
L_y&=\begin{pmatrix}
0&0&i\\0&0&0\\-i&0&0
\end{pmatrix}\\
L_z&=\begin{pmatrix}
0&-i&0\\i&0&0\\0&0&0
\end{pmatrix}
\end{align}
The parameter $\Delta_{\mathrm{so}}$ is the spin-orbit coupling strength. 
In the third line, $K_t$ is the crystalline field in the tangential direction, 
\begin{align}
\bm t(\phi_n)=[-\kappa\sin(\phi_n), p\kappa\cos(\phi_n), |\tau|] .
\end{align}

\begin{figure}
\includegraphics[scale=0.45]{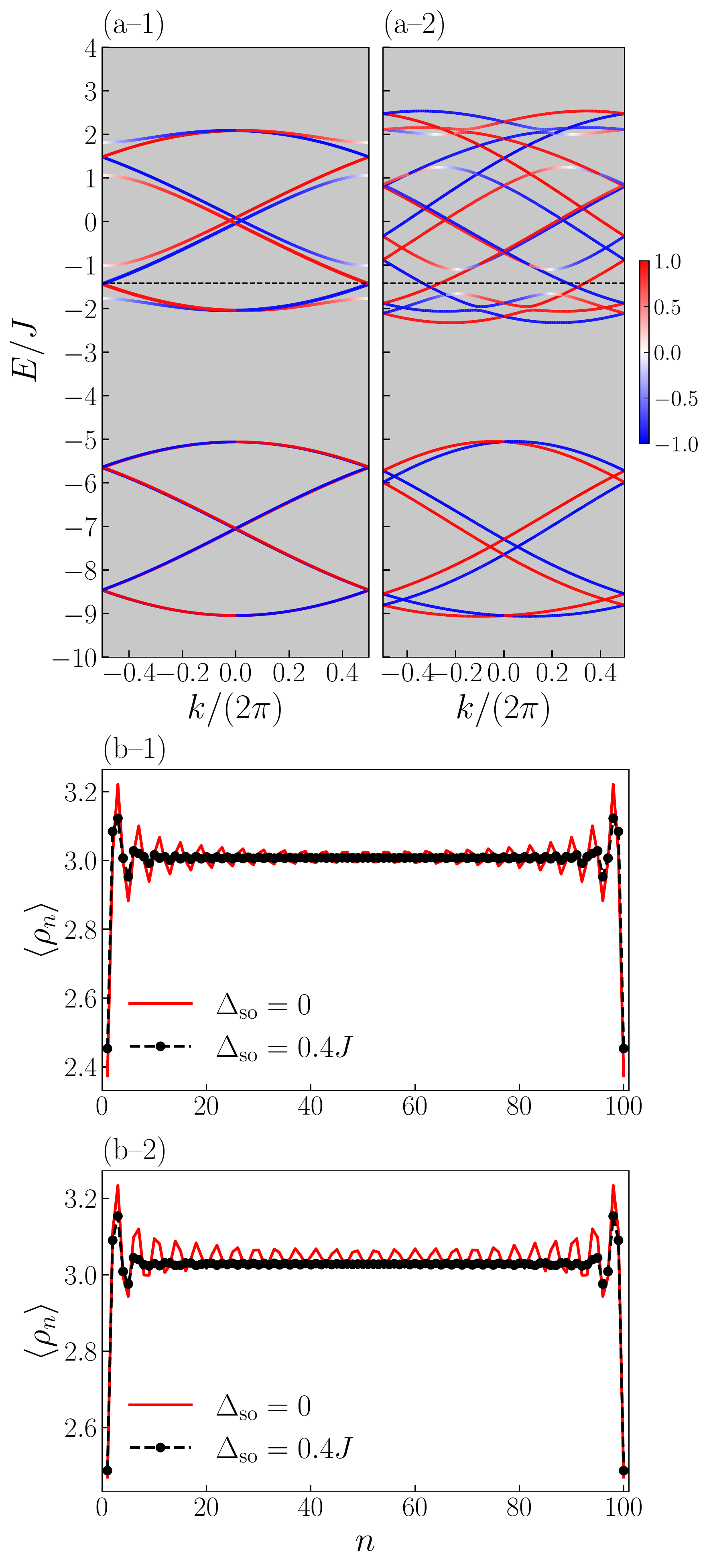}
\caption{\label{dso}
(a-1,2) Band structures for various parameters: $\alpha =1$, $\varphi=\Delta\phi$ and $\tau=0$ for (a-1) and $\alpha =\sqrt{2}$, $\varphi=\Delta\phi/2$ and $\tau=0.48$ for (a-2). 
Other parameters are fixed as $p=1$ and $K_t=7J$.
The colors represent the expectation value of the z-component of spin. (See the color bar.) 
(b-1,2) Integrated charge densities of electrons occupying up to the Fermi level $E_F=-2J \cos \frac{\pi}{N}$ for a finte chain $M =25$ and $N =4$ at $T=0$. 
Parameters of the panels (b-1,2) correspond to those of the panels (a-1,2). 
The Fermi level is indicated by the dashed line in each panel of (a-1,2). 
The marked black dashed lines and red solid lines indicate the local charge density with SOI $\Delta_{\mathrm{so}}=0.4J$ and that without SOI $\Delta_{\mathrm{so}}=0$.}
\end{figure}

Figures \ref{dso} (a-1,2) show the band structures of the infinite chain for various parameters. 
In each panel, we observe two bands separated by $K_t$. 
The panel (a-1) corresponds to the ideal spin-filtering condition~\cite{YU2022,PhysRevB.102.035445}, see Eq. (\ref{Hpm}). 
The helical states are formed in the avoided crossing, the center of which is indicated by the horizontal dashed line at  $E_{0,-}=-2 J \cos \frac{\pi}{N}$. 
The panel (a-2) is the band structure with parameters away from the ideal spin-filtering condition. 
We still observe the two helical states around the horizontal dashed line. 

Figures \ref{dso} (b-1,2) show the local charge density of a finite chain with parameters corresponding to the panels (a-1,2). 
The local charge density is defined as 
\begin{align}
\ev{\rho_n}&=\sum_{E} f(E) \Braket{E|\rho_n|E} , \\
\rho_n &= \Pi_{n} \otimes I_3 \otimes \sigma^{}_0 , 
\end{align}
where the projection operator is a $MN \times MN$ matrix $(\Pi_{n})_{i.j}=\delta_{i,n} \delta_{j,n}$ and $H|E\rangle = E|E\rangle$. 
The electrons are distributed according to the Fermi-Dirac distribution function $f(E)=1/(e^{(E-E_F)/k_{\rm B} T}+1)$, where $k_{\rm B}$ is the Boltzmann constant and $T$ is the temperature. 
Here the temperature is set to be zero, $T=0$. 
The Fermi level is taken at the center of the avoided crossing, 
i.e. $E_F = E_{0,-}$.

In each panel (b-1,2), the solid line and the dashed line are results without and with SOI, respectively. 
We observe the oscillations of charge density in the absence of the SOI.
When the SOI is present, the oscillations in the bulk region are suppressed and the modulations are concentrated at the edges of the molecule. 
We numerically checked that the local charge density is independent of the chirality.

In a single orbital, two electrons can be accommodated, leading to the complete occupancy of the orbital of the lower band. 
The upper band, formed by the mixing of two orbitals, situates the Fermi level at approximately one-fourth of the band width in panels (a-1,2).
Consequently, the average occupancy of electrons is one. 
This explains the reason why the charge density fluctuates around three in the panels (b-1,2).

\section{Analytic results for the local charge density}\label{chargeA}

In this section, we derive an analytic expression of the local charge density.
To make our analysis simpler, we focus on the following conditions:
The torsion is zero, $\tau=0$, the crystalline field along the tangential direction is infinite $K_t\to\infty$, 
$\alpha=1$ and $\varphi=\Delta \phi$ in Eq.~(\ref{mJ}).
Then the original Hamiltonian, Eq.~(\ref{fullH}) can be split into time-reversal symmetric pairs, $H=H_++H_-$, where $H_\pm$ are represented as~\cite{PhysRevB.102.035445}
\begin{multline}
H_\pm= \sum_{n=1}^{MN} \biggl( -J c^\dag_{n+1;\pm} c^{}_{n;\pm}+\mathrm{H.c.} \\
\pm p\Delta_{\mathrm{so}} c^\dagger_{n;\pm}
\begin{bmatrix}
0&e^{-i p\phi_n}\\e^{ip\phi_n}&0
\end{bmatrix}
c^{}_{n;\pm} \biggr). \label{Hpm}
\end{multline}
To obtain this form, we performed a local coordinate transformation at site $n$ as $c_n= e^{i L_z p\phi_n} \tilde{c}_n$. 
The creation operators in the pseudospin space read, 
\begin{align}
c^\dag_{n;+} &=[c^\dag_{n;x\uparrow}\ c^\dag_{n;z\downarrow}] ,\\
c^\dag_{n;-} &=[c^\dag_{n;z\uparrow}\ c^\dag_{n;x\downarrow}] .
\end{align}
$H_+$ hybridizes the up-spin in the $p_x$ orbital and the down-spin in the $p_z$ orbital, and $H_-$ hybridizes the down-spin in the $p_x$ orbital and the up-spin in the $p_z$ orbital in the local coordinate.

\begin{figure*}
\includegraphics[scale=0.34]{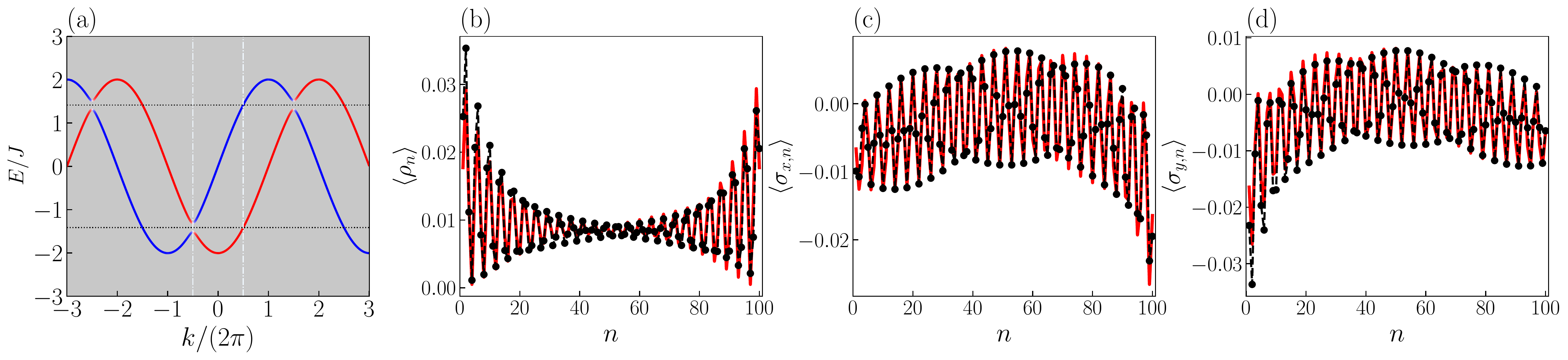}
\caption{\label{charge}
(a) Band structure of $H_+$. 
Upper and lower dashed horizontal lines indicate the center of avoided crossings $E_{0,\pm}= \pm 2 J \cos \frac{\pi}{N}$.
The vertical line indicates the first Brillouin zone.
The local charge density (b) and pseudo-spin $x$ (c) and pseudo-spin $y$ densities of a state at $\delta E=0.035J$, which is close to $E_{0,-}$.
In each panel, the solid red line (dashed black line marked with circles) represents the numerical (analytical) result. 
Parameters: $M=25$, $N=4$, $\Delta_{\rm so}=0.1J$.}
\end{figure*}

In the following, we analyze $H_+$ only.
In order to diagonalize Eq.~(\ref{Hpm}), we take the scattering theory approach~\cite{Matityahu2016}:
We first consider an infinite chain $\hat{H}_+$ and impose a boundary condition.
The eigen energies of $\hat{H}_+$ are
\begin{align}
E_\pm &= E_S(z_{k,p}) \pm p \sqrt{E_A(z_{k,p})^2+\Delta^2_{\rm{so}}} , \label{ez}
\end{align}
where,
\begin{align}
z_{k,p} =& e^{i\frac{k+2\pi p}{N}} ,
\\
E_{S/A}(z_{k,p}) =& \frac{E(z_{k,p}) \pm E(z_{k,p} e^{i \Delta \phi})}{2}  , \label{ESA}
\\
E(z_{k,p}) =&- J \left( z_{k,p}+z^{-1}_{k,p} \right) .
\end{align}
Figure \ref{charge}(a) shows the energy eiganvalue as a function of $k$.
The upper (lower) branch corresponds to $E_+$($E_-$).
If the SOI is absent, the two branches intersect at $E_{0,\pm}=\pm 2J \cos \frac{\pi}{N}$.
The corresponding eigen ket is,
\begin{align}
\ket{z_{k,p};\pm}&=\sum^{\infty}_{n=-\infty} z^{n}_{k,p} \hat{c}^{\dag}_{n;\pm}(z_{k,p})\ket{0} \label{ketkp} ,
\\
\hat{c}_{n;-}^{\dag}(z_{k,p}) =& u(z_{k,p})c^{\dag}_{n;x\uparrow}-v(z_{k,p})e^{i p \phi_n}c^{\dag}_{n;z\downarrow} ,
\\
\hat{c}_{n;+}^{\dag}(z_{k,p}) =& v(z_{k,p})c^{\dag}_{n;x\uparrow}+u(z_{k,p})e^{i p  \phi_n}c^{\dag}_{n;z\downarrow} ,
\end{align}
where,
\begin{align}
u(z_{k,p}) =&\sqrt{\frac{1}{2}\left(1-\frac{p E_A(z_{k,p})}{\sqrt{E_A(z_{k,p})^2+\Delta^2_{\rm{so}}}}\right)}, \label{u}
\\
v(z_{k,p}) =& \sqrt{\frac{1}{2}\left(1+\frac{p E_A(z_{k,p})}{\sqrt{E_A(z_{k,p})^2+\Delta^2_{\rm{so}}}}\right)}. \label{v}
\end{align}
In the following, we focus on the right-handed helix $p=1$. 
We first find the four values of $z_{k,p}$, which are the solutions of $E=E_\pm$, see Eq.~(\ref{ez}):
\begin{align}
z_{s,s'}&=e^{-i\frac{\Delta \phi}{2}}\left(\xi_s-s'\sqrt{\xi^2_s-1}\right), \label{z} \\
\xi_s &= - \frac{E}{2J} \cos\frac{\pi}{N} + s \sqrt{\left[ 1- \left( \frac{E}{2J} \right)^2 \right] \sin^2 \frac{\pi}{N}+\left( \frac{\Delta_{\rm so}}{2J} \right)^2}, \label{xi}
\end{align}
where $s, s'= \pm$. The eigen ket associated with a given energy $E$ is then a linear combination of Eq.~(\ref{ketkp}) with these four values of $z_{k,p}$, 
\begin{align}
\ket{E} = \sum_{s,s'=\pm} a_{s,s'} \ket{z_{s,s'};\sigma_{s}} \, ,  \label{ketE}
\end{align}
where $\sigma_{s}=+$ or $-$. The  probability amplitude at site $n$ is then
\begin{align}
\psi(n)
= \begin{pmatrix}
\psi_\uparrow(n) \\
\psi_\downarrow(n)
\end{pmatrix}
=
\begin{pmatrix}
\langle {n;\uparrow} \ket{E} \\
\langle {n;\downarrow} \ket{E}
\end{pmatrix} .
\end{align}
where we write $|n;\sigma_s \rangle =c^\dagger_{n;o \sigma_s} |0 \rangle$ ($o=p_x,p_z$ in the local coordinate).

The four coefficients $a_{s,s'}$ and the energy $E$ are determined to satisfy the boundary condition consisting of four equations (see Appendix~\ref{coefficients}) and the normalization condition $\langle E \ket{E} = 1$.
The helical states are formed in the energy window,
$| E- E_{0,\pm} |< {\Delta}_{\rm so}$. 
In the following, we consider the states inside the energy window of the lower avoided crossing.
The energy measured from the center of the energy window (dashed line in Fig.~\ref{charge}(a)) is,
$\delta E=E-E_{0,-}$.
For $\delta E>0$, in the leading approximation (see Appendix \ref{details} for detailed calculations),
\begin{align}
z_{-,s'} \approx& e^{-i \frac{\pi}{N}} e^{-i s' \left( \Delta \phi + \delta {k} \right) } , \label{zm} \\
z_{+,s'} \approx& e^{-i \frac{\pi}{N}} e^{-s'/\lambda} , \label{zp}
\end{align}
where $\lambda = 2J \sin \frac{\pi}{N}/\sqrt{\Delta_{\rm so}^2-\delta E^2}$ is the decay length and
$\delta {k}=\delta E/\left(2J \sin \frac{\pi}{N} \right)$ is the wave number.
The former $z_{-,s'}$ corresponds to left ($s'=+$) or right ($s'=-$) going states.
The latter $z_{+,s'}$ corresponds to the evanescent states with the decay length $\lambda$ associated to the avoided crossing.
The eigen ket is the linear combination of $|z_{-,\pm};- \rangle$ and $|z_{+,\pm};+ \rangle$. 
The state close to the center of the avoided crossing, $E \approx E_{0,-}$, reads,
\begin{align}
\psi(n) \propto &
e^{-i \frac{\phi_n}{2} \sigma_z}
\biggl[
\begin{pmatrix}
0\\1
\end{pmatrix}
e^{-i(\Delta\phi+\delta {k})n}
+a
\begin{pmatrix}
1\\0
\end{pmatrix}
e^{i(\Delta\phi+\delta {k})n} \notag\\
&+b
\begin{pmatrix}
1\\i
\end{pmatrix}
e^{-n/\lambda}
+c
\begin{pmatrix}
1\\-i
\end{pmatrix}
e^{n/\lambda}\biggr]
.
\label{wavefunction}
\end{align}
The first and second terms represent the left-going down- and the right-going up-spins.
The remaining two terms are the evanescent states that exhibit exponential decay. 
The coefficients $a$, $b$ and $c$ are determined by the three boundary conditions
(Appendix~\ref{coefficients}).
Figures \ref{charge} (b), (c) and (d) show the local charge density $\Braket{E'|\rho_n|E'} = \Braket{\rho_n}$ and pseudo spin density $\Braket{E'|\sigma_{i,n}|E'} =\Braket{\sigma_{i,n}}$ of a state with the eigenenergy $E'$, which is the closest to $E_{0,-}$. 
In each panel, the analytical result (dashed black line marked with circles) reproduces well the numerical result (solid red line).
In Fig.~\ref{charge} (b), an exponential increase in density is observed around each edge of the molecule.
Figures \ref{charge} (c) and (d) indicate that both $\sigma_x$ and $\sigma_y$ are finite. 
We checked that the $z$-component of spin is nearly zero.

Note that only with the first two terms of Eq.~(\ref{wavefunction}), the left-going and right-going states, it is not possible to fulfill the boundary condition Eqs.~(\ref{bc1}) and ~(\ref{bc2}), since they have opposite spins, see discussions raised in Ref.~\onlinecite{EntinWohlman2021} related to  Ref.~\onlinecite{VarelaPRBR2021}.
The evanescent spins existing in the finite chain can mix the two spins and are necessary to fulfill the boundary condition.
They are not localized edge state, since they hybridize with the left- and right-going states. 

In the above discussion, we only considered $H_+$.
From the subsystem $H_-$, we obtain the same charge density with the opposite spin.
Therefore, in the total system, the spin density vanishes, as expected from the TRS of the original Hamiltonian, Eq.~(\ref{fullH}).

\section{\label{magnet} Effect of the Zeeman field at one edge}

In this section, we turn our attention to the spin state and discuss the enantioselective adsorption to the magnetic substrate.~\cite{Banerjee} 
We assume that the magnetic substrate induced the Zeeman field $B$ at the first site of the atomic chain, 
$H'= H + H_B$: 
\begin{align}
H_B =\mu_{\rm B} B \, \tilde{c}^\dagger_{1}\  {\bm e}_r \cdot \bm\sigma\ \tilde{c}_{1} , \label{magH}
\end{align}
where ${\bm e}_r=(\sin\tilde{\theta}\cos\tilde{\phi}, \sin\tilde{\theta}\sin\tilde{\phi}, \cos\tilde{\theta})$ 
is the direction of the Zeeman field and $\mu_{\rm B}$ is the Bohr magneton. 
The breaking of the TRS induces a finite local spin density at site $n$, which is defined as 
\begin{align}
\ev{\sigma_{i,n}}&=\sum_{E}f(E)\Braket{E|\sigma_{i,n}|E} , \\
\sigma_{i,n}&= \Pi_{n} \otimes I_3 \otimes \sigma_i . 
\end{align}

\begin{figure*}
\includegraphics[scale=0.35]{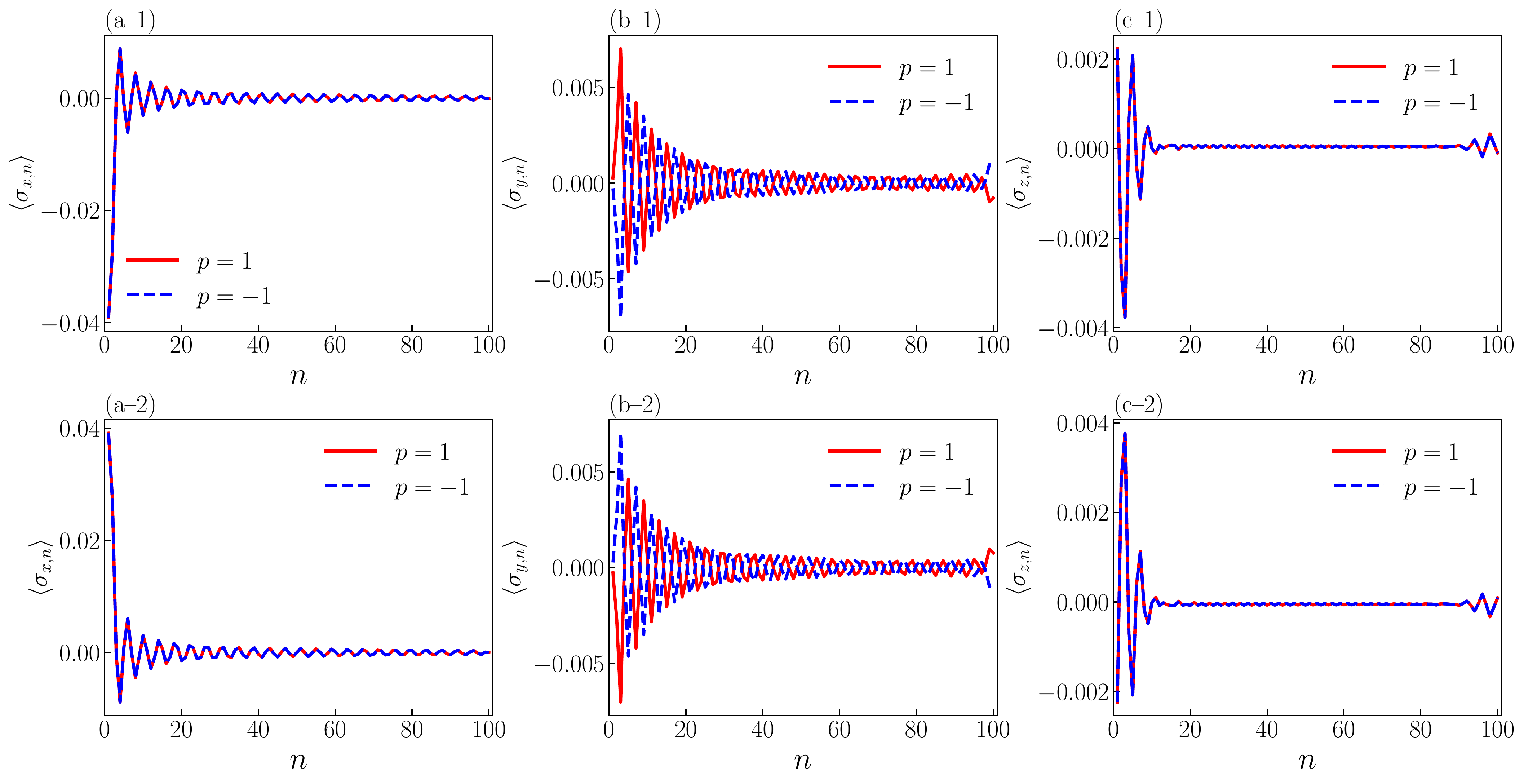}
\caption{\label{magspin}
Local spin densities under the Zeeman field in the $x$ ($-x$) direction, where ${\bm e}_r = \pm (1,0,0)$ (upper/lower panels).
(a), (b), (c) represent the spin density of the $x$, $y$, and $z$ components, respectively.
Parameters: $M=25$, $N=4$, $\Delta_{\rm so}=0.4J$, $E_F=-2J \cos\frac{\pi}{N}$, $\mu_{\rm B} B=0.05J$ and $T=0$.
}
\end{figure*}

The upper/lower panels of Fig.~\ref{magspin} depict the local spin densities when the Zeeman field is applied in the $x$($-x$) direction, i.e. ${\bm e}_r=\pm (1,0,0)$. 
Here, we consider a small Zeeman energy, $\mu_B B \ll \Delta_{\rm so}$, to ensure its impact on the local charge modulations is negligible. 
As observed from the panels, finite spin components are induced around the first site. 

When the direction of the magnetic field is reversed, the induced spin is also reversed, 
$\Braket{\sigma_{i,n}} \to - \Braket{\sigma_{i,n}}$. 
The result is expected from the time-reversal properties of the Hamiltonian. 

On the other hand, when the chirality is reversed, the component perpendicular to the Zeeman field is reversed 
$\Braket{\sigma_{y,n}} \to - \Braket{\sigma_{y,n}}$. 

\begin{figure*}
\includegraphics[scale=0.4]{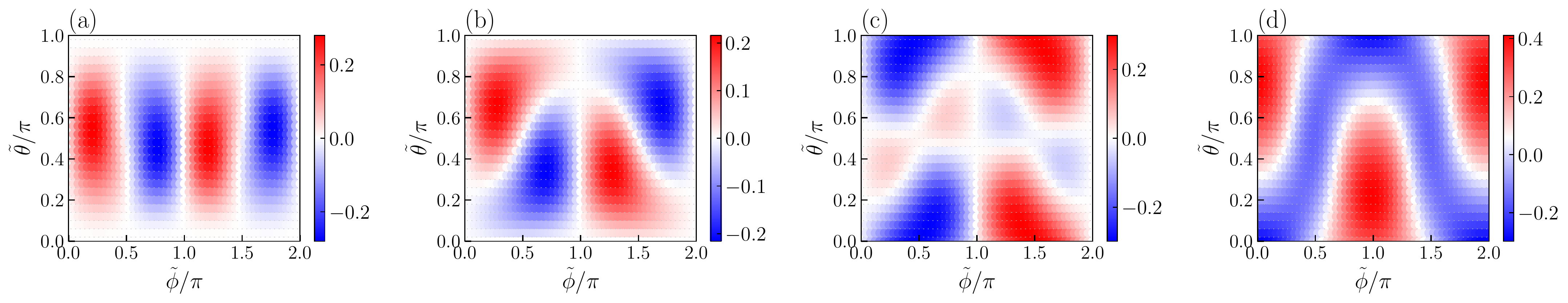}
\caption{\label{defer_spin}
Total energy and total spin changes induced by chirality variation. 
(a) Change in total energy, $\Delta {\mathcal E}/J$.
(b) Change in total spin parallel to the Zeeman field, $\Delta \sigma_r$, and changes perpendicular to the Zeeman field, (c) $\Delta \sigma_{\tilde{\theta}}$ and (d) $\Delta \sigma_{\tilde{\phi}}$. 
Parameters: $M=10$, $N=4$, $\Delta_{\rm so}=0.4J$, $E_F=-2J \cos\frac{\pi}{N}$, $\mu_B B=1J$ and $k_{\rm B}T=0.1 J$.
}
\end{figure*}

The total energy also depends on the chirality.
Figure~\ref{defer_spin} (a) shows the difference in the total energy induced by the change in the chirality, 
\begin{align}
\Delta {\mathcal E} = {\mathcal E}(p=+1)-{\mathcal E}(p=-1) , 
\label{DeE}
\end{align}
for various directions of the Zeeman field. 
Here, we focus on the variation of the total electron energy induced by the SOI:  
\begin{align}
{\mathcal E}(p) =\sum_{E}f(E) \left[ E(p) - \left. E(p) \right|_{ \Delta_{\rm so}= 0} \right], 
\end{align}
The typical value of the hopping energy for a DNA molecule is $J \approx 25$meV.~\cite{Gutierrez2012}
We set $\Delta_{\rm so} =0.4 J \approx 10$meV, which is approximately the intra-atomic spin-orbit coupling energy in carbon nanotubes.~\cite{Huertas-Hernando2006} 
In our numerical calculations, the Zeeman energy is set to be compatible with the hopping energy $\mu_{\rm B} B=J$. 
In this scenario, the maximum energy change can reach up to $\Delta {\mathcal E}_{\rm max} \approx 0.2 J \approx 50$K,
which is not negligible and is larger than the temperature $k_{\rm B} T=0.1 J \approx 25$K adopted in our numerical calculations. 

Panels (b-d) in Fig.~\ref{defer_spin} depict the alteration in the total spin, caused by the changes of the chirality: 
\begin{align}
\Delta \sigma_i &= \sigma_i(p=+1)-\sigma_i(p=-1) , \label{DeM} \\
\sigma_i(p) &= \sum_{n=1}^{MN}\ev{\sigma_{i,n}(p)} . 
\end{align}
Panel (b) illustrates the component parallel to the magnetic field direction, $\Delta \sigma_r = \Delta {\bm \sigma} \cdot {\bm e}_r$.
Panels (c) and (d) represent components perpendicular to the magnetic field direction, 
$\Delta \sigma_{\tilde{\theta}} = \Delta {\bm \sigma} \cdot {\bm e}_{\tilde{\theta}}$. 
and
$\Delta \sigma_{\tilde{\phi}}  = \Delta {\bm \sigma} \cdot {\bm e}_{\tilde{\phi}}$, 
where 
${\bm e}_{\tilde{\theta}} = \partial_{ \tilde{\theta}} {\bm e}_r/|\partial_{ \tilde{\theta}} {\bm e}_r|$
and
${\bm e}_{\tilde{\phi}} = \partial_{ \tilde{\phi}} {\bm e}_r/|\partial_{ \tilde{\phi}} {\bm e}_r|$. 
In each panel, we observe a maximum of $0.2 \hbar$ to $0.4 \hbar$ spin angular momentum.

\begin{figure*}
\includegraphics[scale=0.4]{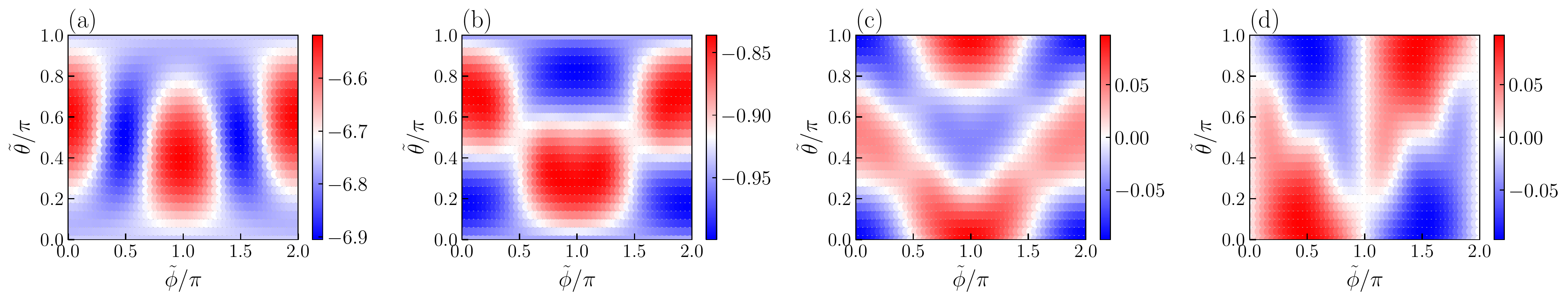}
\caption{\label{average_spin}
Averages for different chiralities. 
(a) Average of the total energy, $\overline{ {\mathcal E} }/J$.
(b) Average of the total spin parallel to the Zeeman field, $\overline{\sigma}_r$, and the averages perpendicular to the Zeeman field, 
(c) $\overline{\sigma}_{\tilde{\theta}}$ and (d) $\overline{\sigma}_{\tilde{\phi}}$. 
The parameters are the same as those in Fig.~\ref{defer_spin}. 
}
\end{figure*}

Figure \ref{average_spin} represents the averages for different chiralities: 
\begin{align}
\overline{ {\mathcal E} }&= \Big[ {\mathcal E}(p=+1)+{\mathcal E}(p=-1) \Bigr]/2 , \label{aveE} \\
\overline{\sigma}_i&=\Big[ \sigma_i(p=+1)+\sigma_i(p=-1) \Bigr]/2.\label{aveM}
\end{align}
Comparing Figs.~\ref{defer_spin} and \ref{average_spin}, the maximum of $|\Delta \sigma_r|$ is approximately 20\% of the maximum of $|\overline{\sigma}_r|$. 
On the other hand, the maximum of $|\Delta \sigma_{\tilde{\theta} (\tilde{\phi}) }|$ is about 4 to 5 times larger than the maximum of $|\overline{\sigma}_{\tilde{\theta} (\tilde{\phi}) }|$.
Therefore, the chirality change induces primarily the change in the spin perpendicular to the magnetic field direction. 
Figure~\ref{average_spin} (a) indicates that the SOI induces the total energy variation about $7J$.
By comparing it with Figs.~\ref{defer_spin}(a) ,  we conclude that only a small portion of the total energy induced by the SOI depends on chirality. 

In the above results, we took a relatively large Zeeman field, $\mu_{\rm B} B \approx 25$meV. 
The exchange field as large as $\mu_{\rm B} B \approx 15$meV has been observed for a spin in the C$_{60}$ molecular quantum dot tunnel-coupled to the Ni electrodes.~\cite{Pasupathy2004} 
This exchange interaction results from the spin-dependent quantum charge fluctuations between the magnetic leads and the molecule.~\cite{Martinek2003,Utsumi2005}
The situation seems not to agree with the spin-dependent dispersion force scenario.~\cite{Naaman2020,Kumar2017}
Note that such a large exchange interaction was observed when the tunnel coupling to the ferromagnetic lead is sufficiently strong, leading to the realization of the Kondo state.~\cite{Pasupathy2004}
However, we are uncertain whether a similar strong tunnel coupling can be attained for a mass of chiral molecules adsorbed on a magnetic substrate. 
Our theoretical model does not exclude other possibilities, such as spin-dependent dispersion forces.

\section{\label{conclusion}Conclusion}

We discuss the electronic and spin states of the $p$-orbital helical atomic chain with the intra-atomic spin-orbit interaction. 
For the infinite length and specific parameters, this model has two avoided crossings and in each energy window, there exist two helical states.
We demonstrate that when the Fermi energy is in this energy window, the charge modulations concentrate at the edges.
Analytically solving the wave function, we found that this happens because of the evanescent states located at the edges. 
Charge modulations concentrated at the edges become spin-polarized when a magnetic field is applied. 
The Zeeman field at one edge, which simulates the effect of the magnetic substrate in enantioselective adsorption experiments, induces a finite chirality-dependent spin polarization. 
Figure \ref{magspin} shows that the helicity affects only the spin component which is perpendicular to the axis of the molecule and to the direction of the magnetic field, which is generated by the polarized substrate. 
This effect occurs only when both spin-orbit interaction and a magnetic field are present.
The spin component oscillates and decays along the helix, but has opposite signs for the two helicities. It would be interesting to test these predictions, by some local measurements of the magnetization on the molecule. 
Although it decays, it still maintains a finite value even at the other end of the molecule.
Figure \ref{defer_spin} demonstrates the induced spin varies depending on the direction of the Zeeman field: the primary chirality dependent component appears perpendicular to the Zeeman field. 
This chirality-dependent spin polarization also causes the energy differences between different chiralities, which provides an insight into the enantioselectivity in CISS.
To explain the experiment, we speculate that, in addition to the ferromagnetic substrate, further effects such as Coulomb interaction and lattice vibrations~\cite{KatoPRB2022,Fransson2020,Michaeli,Plenio2022,Plenio2023} need to be accounted for.

\begin{acknowledgments}
This work was supported by Marubun Research Promotion Foundation, Spintronics Research Network of Japan (Spin-RNJ), JSPS KAKENHI Grants No. 18KK0385,No. 20H01827 and No. 20H02562.
\end{acknowledgments}

\appendix

\section{Detailed derivations}
\label{details}

Equation~(\ref{xi}) can be expanded up to the accuracy of $\delta E^2$ and $\Delta_{\rm so}^2$ as,
\begin{align}
\xi_s \approx 1+(s-1) \left( \sin^2 \frac{\pi}{N} + \frac{\delta E}{2J} \cos \frac{\pi}{N} \right) + \frac{s}{2\lambda^2} . \label{xi_expand}
\end{align}

For $s=-$, by substituting Eq.~(\ref{xi_expand}) into Eq.~(\ref{z}), we obtain Eq.~(\ref{zm}) up to $\delta E$ accuracy.
Then by using Eq.~(\ref{ESA}), we obtain
\begin{align}
E_A(z_{-,s'}) \approx s'2J\sin\frac{\pi}{N}\sin 2\zeta.
\end{align}
Where $\zeta=\sqrt{\sin^2 \left(\frac{\pi}{N}\right) + \left(\frac{\delta E}{2J}\right) \cos \frac{\pi}{N}}$\ .
Considering the conditions, $|E_A(z_{-,s'})| \gg \Delta_{\rm so}$
, Eqs.~(\ref{u}) and (\ref{v}) become,
\begin{align}
u(z_{-,s'})&\approx\sqrt{\frac{1}{2}\left(1-s'\right)}=\delta_{s',-}, \\
v(z_{-,s'})&\approx\sqrt{\frac{1}{2}\left(1+s'\right)}=\delta_{s',+}.
\end{align}
Here we used $p=1$.
Then the wave functions of left and right going sates are,
\begin{align}
\begin{pmatrix}
\langle n; \uparrow |z_{-,s'};+ \rangle \\
\langle n;\downarrow |z_{-,s'};+ \rangle
\end{pmatrix}
=&
z_{-,s'}^n
\begin{pmatrix}
v(z_{-,s'}) \\
u(z_{-,s'})e^{i \phi_n}
\end{pmatrix}
\notag \\
& \approx
e^{-i \frac{\phi_n}{2} \sigma_z }
e^{-i  s' \left( \phi_n+\delta {k} n \right) }
\begin{pmatrix}
\delta_{s',+} \\
\delta_{s',-}
\end{pmatrix} ,
\\
\begin{pmatrix}
\langle {n; \uparrow} |z_{-,s'};- \rangle \\
\langle {n; \downarrow} |z_{-,s'};- \rangle
\end{pmatrix}
=&
z_{+,s'}^n
\begin{pmatrix}
u(z_{-,s'}) \\
-v(z_{-,s'})e^{i \phi_n}
\end{pmatrix}
\notag \\
&\approx
e^{-i  \frac{\phi_n}{2} \sigma_z }
e^{-i  s' \left( \phi_n+ \delta {k} n \right) }
\begin{pmatrix}
  \delta_{s',-} \\
 -\delta_{s',+}
\end{pmatrix}
.
\end{align}
Similarly for $s=+$, we obtain, Eq.~(\ref{zp}) up to $1/\lambda$ accuracy
and
\begin{align}
E_A(z_{+,s'})=-i 2 J \sin \frac{\pi}{N} \sinh (s'/\lambda)
\approx -i s '\sqrt{\Delta^2_{\rm{so}}-\delta E^2},
\end{align}
Then the wave functions of evanescent states are,
\begin{align}
\begin{pmatrix}
\langle {n; \uparrow} |z_{+,s'};+ \rangle \\
\langle {n; \downarrow} |z_{+,s'};+ \rangle
\end{pmatrix}
\approx &
e^{-i\frac{\phi_n}{2} \sigma_z} e^{-s' n/\lambda}
\begin{pmatrix}
\sqrt{\frac{1}{2}\left(1 - i s' \frac{\sqrt{\Delta_{so}^2-\delta E^2}}{|\delta E|} \right)}\\
\sqrt{\frac{1}{2}\left(1 + i s' \frac{\sqrt{\Delta_{so}^2-\delta E^2}}{|\delta E|} \right)}
\end{pmatrix}
,
\\
\begin{pmatrix}
\langle {n; \uparrow} |z_{+,s'};- \rangle \\
\langle {n; \downarrow} |z_{+,s'};- \rangle
\end{pmatrix}
\approx &
e^{-i\frac{\phi_n}{2} \sigma_z} e^{-s' n/\lambda}
\begin{pmatrix}
\sqrt{\frac{1}{2}\left(1 + i s' \frac{\sqrt{\Delta_{so}^2-\delta E^2}}{|\delta E|} \right)}\\
-\sqrt{\frac{1}{2}\left(1 - i s' \frac{\sqrt{\Delta_{so}^2\delta E^2}}{|\delta E|} \right)}
\end{pmatrix}.
\end{align}
Close to the center of the lower avoided crossing, $|\delta E| \ll \Delta_{\rm so}$, they are
\begin{align}
\begin{pmatrix}
\langle {n; \uparrow} |z_{+,s'};+ \rangle \\
\langle {n; \downarrow} |z_{+,s'};+ \rangle
\end{pmatrix}
\approx&
e^{-i\frac{\phi_n}{2} \sigma_z}
e^{-s' n/\lambda}
\sqrt{ \frac{\Delta_{so}}{2|\delta E|} }
\begin{pmatrix}
  e^{-is' \pi/4} \\
  e^{is' \pi/4}
\end{pmatrix}
,
\\
\begin{pmatrix}
\langle {n; \uparrow} |z_{+,s'};- \rangle \\
\langle {n; \downarrow} |z_{+,s'};- \rangle
\end{pmatrix}
\approx&
e^{-i\frac{\phi_n}{2} \sigma_z} e^{-s' n/\lambda}
\sqrt{ \frac{\Delta_{so}}{2 |\delta E|} }
\begin{pmatrix}
  e^{is' \pi/4} \\
-e^{-is' \pi/4}
\end{pmatrix}.
\end{align}
The linear combination,
\begin{align}
|E \rangle
=&
-|z_{-,+};- \rangle
+a |z_{-,-};- \rangle
\notag \\
&+b e^{i \pi/4}
\sqrt{ \frac{2 |\delta E|}{\Delta_{so}} }
|z_{+,+};+ \rangle
+c e^{-i \pi/4}
\sqrt{ \frac{2 |\delta E|}{\Delta_{so}} }
|z_{+,-};+ \rangle
,
\end{align}
results in Eq.~(\ref{wavefunction}).

\section{Coefficients of wave function}
\label{coefficients}

To determine the coefficients, we consider the following boundary condition,
\begin{align}
\ell_+(0) = \ell_+(MN+1)&=0,\label{bc1}\\
\ell_-(0) + \ell_-(MN+1) &= 0 \, ,\label{bc2}
\end{align}
where
$\ell_\pm(n)=\psi_\uparrow(n) \pm \psi_\downarrow(n)$. 
The 3 coefficients in Eq.~(\ref{wavefunction}) are,
\begin{align}
a=&[ (i-1)(i+\alpha^2)\beta+i(\alpha^2-1)\gamma+2\alpha\beta\gamma \nonumber \\ &+\beta^2(\gamma+\alpha(\alpha\gamma-2))]/\eta, \\
b=&-\frac{\alpha(\beta+\gamma)(-\beta-i\gamma+\alpha(i+\beta\gamma))}{\eta} , \\
c=&\frac{(\beta+\gamma)(i+\alpha(\beta-i\gamma)-\beta\gamma)}{\eta}.
\end{align}
where
\begin{align}
\alpha=&e^{(MN+1)/\lambda}, \\
\beta=&e^{i(MN+1)\Delta\phi/2}, \\
\gamma=&e^{i(MN+1)(\Delta\phi + \delta k)}, \\
\eta=&\gamma[-1-2\alpha(\beta-\gamma)-i\beta(\beta+(1+i)\gamma)\notag\\
&+\alpha^2(-1+i\beta^2+(1+i)\beta\gamma)].
\end{align}
To obtain Fig.~\ref{charge}, 
we calculate (here, we denote $\rho$ as $\sigma_0$)
\begin{align}
\frac{\psi(n)^\dagger \sigma_i \psi(n)}{\sum_{n=1}^{MN} \psi(n)^\dagger \psi(n)}\hspace{5mm}
(i=0,x,y).
\end{align}

\nocite{*}
\bibliography{JCP}

\end{document}